\documentclass[amsmath,superscriptaddress,showpacs,aps,prl,twocolumn]{revtex4-1}
\usepackage[colorlinks,linkcolor=blue,anchorcolor=blue,citecolor=blue,urlcolor=blue]{hyperref}
\usepackage{amsmath}
\usepackage{amssymb}
\usepackage{graphicx}
\usepackage{color}
\usepackage{bm}

\begin{document}
\title{Quantitative Determination of the Confinement and Deconfinement of spinons in the anomalous spectra of Antiferromagnets via the Entanglement Entropy}

\author{Zhao-Yang Dong}
\affiliation{Department of Applied Physics, Nanjing University of Science and Technology, Nanjing 210094, China.}
\author{Wei Wang}
\email[]{wwang@njupt.edu.cn}
\affiliation{School of Science, Nanjing University of Posts and Telecommunications, Nanjing 210023, China}
\author{Zhao-Long Gu}
\email[]{waltergu1989@gmail.com}
\affiliation{National Laboratory of Solid State Microstructures and Department of Physics, Nanjing University, Nanjing 210093, China}
\author{Jian-Xin Li}
\email[]{jxli@nju.edu.cn}
\affiliation{National Laboratory of Solid State Microstructures and Department of Physics, Nanjing University, Nanjing 210093, China}
\affiliation{Collaborative Innovation Center of Advanced Microstructures, Nanjing University, Nanjing 210093, China}
\date{\today}

\begin{abstract}

\par We introduce an entanglement entropy analysis to quantitatively identify the confinement and deconfinement of the spinons in the spin excitations of quantum magnets. Our proposal is implemented by the parton construction of a honeycomb-lattice antiferromagnet exhibiting high-energy anomalous spectra. To obtain the quasiparticles of spin excitations for entanglement entropy calculations, we develop an effective Hamiltonian using the random phase approximation. We elaborate quantitatively the deconfinement-to-confinement transition of spinons in the anomalous spectra with the increase of the Hubbard interaction, indicating the avoided fractionalization of magnons in the strong interaction regime. Meanwhile, the Higgs mode at the $\Gamma^{\prime}$ point is fractionalized into four degenerate spinon pairs, although it appears as a sharp well-defined peak in the spectra. Our work extends our understanding of the deconfinement of the spinon and its coexistence with the magnon in quantum magnets.

\end{abstract}

\maketitle

\par The nature of the ground states and their respective elementary spin excitations resides at the center of the field of quantum magnetism. For quantum magnets with strong frustrations, quantum fluctuations can suppress the formation of local magnetic orders, thus, exotic spin liquid phases can be stabilized, with deconfined spinons as the elementary excitations \cite{Balents2010}. On the other hand, for quantum magnets in the weak-frustration or frustration-free regime, the system hosts long-range magnetic orders, and the low-energy physics can be captured quite well by the semiclassical approach, i.e., the spin wave theory (SWT), where the elementary excitations are well-defined magnons. However, this does not imply that the quantum fluctuations are negligible for ordered quantum magnets. In fact, the semiclassical ordered states are not their exact ground states in the presence of fluctuations which usually reduce the magnitude of the local magnetic order significantly \cite{Reger_PRB1988, Hamer_PRB1992}. The ground state of an ordered quantum magnet may be close to a spin liquid state, such as the resonating-valence-bond (RVB) state \cite{Anderson_PRL1987}.

\par Recent studies on the spin excitation spectra further reveal the quantum nature of the magnetically ordered quantum magnets. Compared to the SWT predictions, inelastic neutron scattering experiments on the square-lattice antiferromagnetic (AF) compounds \cite{Christensen2007, DallaPiazza2015, PhysRevLett.105.247001} uncovered a remarkable anomaly at $(\pi, 0)$ of the Brillouin zone (BZ) in the high-energy spectra, where the energies of the transverse spin excitations are renormalized downward to form a roton-like minimum. Meanwhile, the sharp well-defined magnonic modes disappear around this point with a dim broad continuum left. Similar anomalous behaviors were also observed in other AF compounds on different two-dimensional lattices experimentally \cite{PhysRevLett.109.267206, PhysRevLett.110.267201, PhysRevLett.116.087201, Ito2017, Kamiya2018, PhysRevLett.108.057205, PhysRevB.102.014427, Wessler2020, PhysRevB.100.180406, Sala2021}. Such anomalies indicate that additional spin fluctuations, such as those in the adjacent spin liquid phase, would survive in the AF ordered ground state. To appropriately capture the physics of the spin excitations at all energies, it appears more suitable to describe the ordered quantum magnets in terms of spinons \cite{PhysRevB.65.165113, Zheng2006, PhysRevX.8.011012}. In this scenario, the magnon is a bound state of two spinons lying in the gap of the two-spinon continuum. In fact, based on the variational RVB ansatz with magnetic orders, the anomaly in the high energy spin excitation spectra is attributed to the deconfinement of spin-1/2 spinons \cite{DallaPiazza2015, PhysRevX.9.031026, Zhang2020, Yu2018, Ferrari2018, Ferrari2020}. This scenario is supported by various numerical studies \cite{Hsu1990, Syljuasen2000, Ho2001, Zheng2006, PhysRevLett.110.217213, DallaPiazza2015, Ghioldi2016, Shao2017, PhysRevLett.119.157203, PhysRevX.9.031026, Yu2018, Ferrari2018, Ferrari2020, Zhang2020}, including the unbiased large-scale quantum Monte Carlo simulations \cite{Shao2017}.

\par In this spinon-based interpretation, the main evidence in favor of the deconfinement of spinons in the anomalous spectra of ordered magnets is that the continuum can be continuously evolved into that appeared in a neighboring disordered state in the phase diagram by varying some model parameters, which indeed consists of deconfined particle-hole spinon pairs \cite{Shao2017, Yu2018, PhysRevX.9.031026}. However, with a fixed given set of parameters, the deconfinement can only be phenomenologically determined by the evolution of the spectral line shape from a sharp Lorentzian peak to a broad continuum \cite{Shao2017, Yu2018, PhysRevX.9.031026, Zhang2020}. In nature, it still remains elusive to distinguish spinons from magnons when they are coupled in the anomalous spin spectra. In this letter, we propose the entanglement entropy (EE) of a spin excitation as the hallmark to identify the deconfinement of spinons in the excitation spectra of quantum magnets unambiguously and quantitatively.

\par Our proposal is implemented on a honeycomb-lattice antiferromagnet in the weak-frustration regime exhibiting high-energy anomalous spectra, which is relevant to the recent experiments on YbCl$_3$ and YbBr$_3$ \cite{PhysRevB.102.014427, Wessler2020, Sala2021}. To obtain the quasiparticles of spin excitations for EE calculations, we develop an effective Hamiltonian using the random phase approximation (RPA) based on a variational SU($2$) RVB ansatz with local magnetic orders. We elaborate quantitatively the deconfinement-to-confinement transition of spinons in the anomalous spectra with the increase of the Hubbard interaction, indicating the avoided fractionalization of magnons in the strong interaction regime. Meanwhile, the Higgs mode at the $\Gamma^{\prime}$ point is fractionalized into four degenerate spinon pairs despite it appears as a sharp well-defined quasiparticle peak from the spectral perspective. Our work reveals
the insufficiency of the spectral perspective and provides an quantitatively accurate judgement on the confinement
and deconfinement of spinons.


\begin{figure}
\centering
\includegraphics[width=0.38\textwidth]{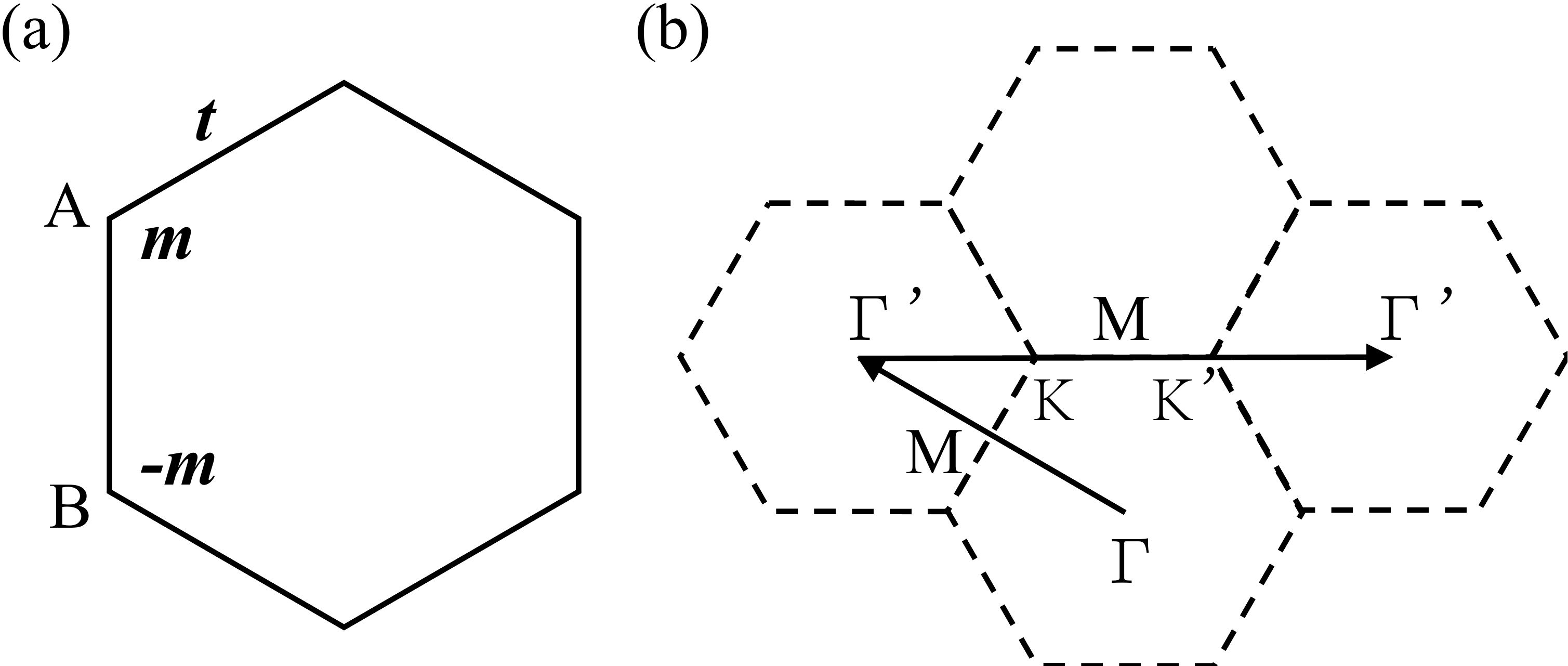}
\caption{(a) Illustration of Eq.~(\ref{Hmean}) in the honeycomb lattice. (b) High symmetry points in the Brillouin zone.}
\label{IPMC}
\end{figure}

\par In the honeycomb lattice, several structures could be close to the AF phase \cite{Meng2010, PhysRevB.82.024419, PhysRevB.84.024420} with the introduction of frustrations, and it has been shown by the mean-field theory that the most energetically favorite one is the SU($2$) RVB ansatz \cite{SM}. Therefore, we adopt this ansatz with the AF order [see Fig.~\ref{IPMC}(a)] as the starting point,
\begin{equation}
  H_{\rm MF}=t\sum_{\langle ij\rangle\alpha}f_{i\alpha}^\dagger f_{j\alpha}+\sum_{i\alpha\beta}m_i f_{i\alpha}^\dagger \sigma^z_{\alpha\beta}f_{i\beta}.
  \label{Hmean}
\end{equation}
Here $t=-0.158J_1$ is the SU($2$) RVB hopping parameter and $m_A=-m_B=m=0.191J_1$ is the AF order parameter obtained by a self-consistent mean-field solution \cite{SM}. The SU($2$) RVB ansatz implies the fluctuations of the plaquette valence bonds \cite{PhysRevLett.84.3173, Mosadeq2011, Bishop2012, PhysRevB.89.094413, PhysRevLett.110.127205}. It is noted that the spiral spin orders \cite{PhysRevB.81.214419, PhysRevB.89.094413} are not included, since we will focus on the spin-liquid-like fluctuations surviving in the AF phase. In this case, Eq.~(\ref{Hmean}) can be diagonalized with $f_{\mathbf{k}\alpha\sigma}=\sum_{\beta\sigma'}u_{\alpha\beta}^{\sigma\sigma'}(\mathbf{k})d_{\mathbf{k}\beta\sigma'}$ with $u_{\alpha\beta}^{\sigma\sigma'}(\mathbf{k})$ being the matrix elements of $U^{\sigma\sigma'}(\mathbf{k})=\delta^{\sigma\sigma'}U^\sigma(\mathbf{k})$:
\begin{eqnarray*}
  U^\uparrow(\mathbf{k})&=&\frac{1}{D}\left(
                  \begin{array}{cc}
                    \varepsilon(\mathbf{k})+m & t\gamma^\ast(\mathbf{k}) \\
                    t\gamma(\mathbf{k}) & -(\varepsilon(\mathbf{k})+m)\\
                  \end{array}
                \right),\\
  U^\downarrow(\mathbf{k})&=&\frac{1}{D}\left(
                  \begin{array}{cc}
                    t\gamma^\ast(\mathbf{k}) & -(\varepsilon(\mathbf{k})+m)\\
                    \varepsilon(\mathbf{k})+m & t\gamma(\mathbf{k})\\
                  \end{array}
                \right),
\end{eqnarray*}
where $D=\sqrt{2\varepsilon(\mathbf{k})(\varepsilon(\mathbf{k})+m)}$, $\gamma(\mathbf{k})=(1+e^{i \mathbf{k}\cdot r_1}+e^{i \mathbf{k}\cdot r_2})$ and $\pm\varepsilon(\mathbf{k})=\pm \sqrt{|t\gamma(\mathbf{k})|^2+m^2}$ the spinon dispersion.

\par To get the spin dynamics beyond the mean-field level, RPA can be applied \cite{Ho2001, Zhang2020}. To represent the fluctuations in the spin density channel, we introduce the Heisenberg interaction $J_{\rm eff}\mathbf{S_i}\cdot\mathbf{S_j}$ between the spinons on the nearest-neighbor sites. We also take into account the Hubbard interaction $U_{\rm eff}n_{i\uparrow}n_{i\downarrow}$ to simulate partially the effect of the no-double-occupancy constraint on the spinons. They satisfy $J_{\rm eff}+2/3U_{\rm eff}=0.4J_1$ in the self-consistent calculation. In the remaining of this paper, we take the proportion of the Hubbard interaction denoted by $a$ as the variable to show the evolution of the spin dynamics, i.e., we take such parameterizations: $U_{\rm eff}=a0.6J_1,~J_{\rm eff}=(1-a)0.4J_1$. It is noted that both $U_{\rm eff}$ and $J_{\rm eff}$ should be understood as phenomenological parameters in the spinon description, which are introduced to mimic the gauge fluctuation over the mean-field ground state and induce a static magnetic order in the system simultaneously \cite{Zhang2020}.

\par With these interactions, the dynamical spin susceptibility $\boldsymbol{\chi}$ can be obtained by the RPA correction on the bare $\boldsymbol{\chi^0}$ of Eq.~(\ref{Hmean}),
\begin{equation}\label{rpa}
  \boldsymbol{\chi}(\mathbf{q},\omega)=\frac{\boldsymbol{\chi^0}(\mathbf{q},\omega)}{\mathbf{I}+\mathbf{V(q)}\boldsymbol{\chi^0}(\mathbf{q},\omega)},
\end{equation}
where $\mathbf{V(q)}=-U_{\rm eff}\mathbf{I}+\mathbf{J(q)}$. $\mathbf{J(q)}$ is the Fourier transform of the Heisenberg interaction,
\begin{equation*}
  \mathbf{J(q)} = \left(
          \begin{array}{cc}
            0 & J_{\rm eff}\mathbf{I_{3\times3}}\gamma(\mathbf{q}) \\
            J_{\rm eff}\mathbf{I_{3\times3}}\gamma^\ast(\mathbf{q}) & 0 \\
          \end{array}
        \right).
\end{equation*}
The intensities of the transverse and longitudinal spin dynamics is related to the spectral functions,
\begin{eqnarray}
  I^{-+}(\mathbf{q},\omega)&=&\frac{1}{\pi}{\rm Im}[\chi^{-+}(\mathbf{q},\omega+i0^+)],\label{sft}\\
  I^{zz}(\mathbf{q},\omega)&=&\frac{1}{\pi}{\rm Im}[\chi^{zz}(\mathbf{q},\omega+i0^+)].\label{sfl}
\end{eqnarray}

\par In the zero-temperature RPA, the spinon particle-hole operators can be approximately regarded as quasibosonic operators $b^\dagger_{\mathbf{k}\mathbf{q}\sigma\sigma'}\simeq d_{\mathbf{k}-\mathbf{q}1\sigma}^\dagger d_{\mathbf{k}2\sigma'}$. Thus, we can develop an effective Hamiltonian about the spin excitations from RPA \cite{SM},
\begin{eqnarray*}
  &&H_{\rm eff}(\mathbf{q})=\sum_{\mathbf{k}\sigma\sigma'}[(\varepsilon(\mathbf{k}-\mathbf{q})+\varepsilon(\mathbf{k}))b^\dagger_{\mathbf{k}\mathbf{q}\sigma\sigma'}b_{\mathbf{k}\mathbf{q}\sigma\sigma'}\\
 && +(\varepsilon(\mathbf{k}+\mathbf{q})+\varepsilon(\mathbf{k}))b_{\mathbf{k}\mathbf{\bar{q}}\sigma\sigma'}b^\dagger_{\mathbf{k}\mathbf{\bar{q}}\sigma\sigma'}]
 +\Phi^\dagger_\mathbf{q} \bar{\bar{U}}_\mathbf{q}^\dagger V_\mathbf{q} \bar{\bar{U}}_\mathbf{q}\Phi_\mathbf{q},
\end{eqnarray*}
where $\Phi^\dagger_\mathbf{q}=(\cdots b^\dagger_{\mathbf{k}\mathbf{q}\sigma\sigma'}\cdots b_{\mathbf{k}\mathbf{\bar{q}}\sigma\sigma'}\cdots)$,  $\bar{\bar{U}}_\mathbf{q}=\mathbf{I}_{N_\mathbf{k}\times N_{\mathbf{k}}}\otimes U^\ast(\mathbf{k})\otimes U(\mathbf{k}-\mathbf{q}) $, and $V_\mathbf{q}$ is the coefficient matrix of the Fourier transform of the Hubbard and Heisenberg terms $[V_\mathbf{q}]_{\mathbf{k}\alpha\beta;\mathbf{k'}\alpha'\beta'}f^\dagger_{\mathbf{k}-\mathbf{q}\alpha}f_{\mathbf{k}\beta}f^\dagger_{\mathbf{k'}\alpha'}f_{\mathbf{k'}-\mathbf{q}\beta'}$. Then the $i$th quasiparticle $\widehat{\Psi}_i(\mathbf{q})$ of the spin excitations is solved from $H_{\rm eff}$ as a linear combination of the products of such particle-hole spinon pairs:
\begin{equation}\label{quasip}
\widehat{\Psi}_i(\mathbf{q})=\sum_{\mathbf{k}\gamma\sigma\sigma'}
 \psi_{i\mathbf{k\gamma\sigma\sigma'}}(\mathbf{q}) d_{\mathbf{k}-\mathbf{q}\gamma\sigma}^\dagger\otimes d_{\mathbf{k}\overline{\gamma}\sigma'},
\end{equation}
where $\gamma$ denotes the conduction or valence band.

\par Since Eq.~(\ref{quasip}) is the Schmidt decomposition of particle and hole spinons, with respect to this bipartition, EEs can be defined to analyze the nature of spin excitations \cite{Dong2020}. The EEs of magnons, for which the particle-hole pairs are confined to form bound states, are logarithmically divergent in the thermodynamic limit, while, the EEs of the modes in the spinon continuum, which can be decomposed into relatively free particle and hole spinons, converge to constants. The EEs, as the measure to distinguish between the confinement and deconfinement of spin excitations, can quantitatively determine whether the fractionalization of magnons into spinons occur. Though the coefficient $\psi_{i\mathbf{k\sigma\sigma'}}(\mathbf{q})$ is not normalized, it can be viewed as a vector, whose $\alpha$-norm is well defined, thus, the R\'{e}nyi entropy reads,
\begin{equation}\label{EE}
  S_i(\mathbf{q})=\frac{1}{1-\alpha}{\rm ln}(\sum_{\mathbf{k}\gamma\sigma\sigma'} |\psi_{i\mathbf{k\gamma\sigma\sigma'}}(\mathbf{q})|^{2\alpha}).
\end{equation}

\begin{figure}
\centering
\includegraphics[width=0.45\textwidth]{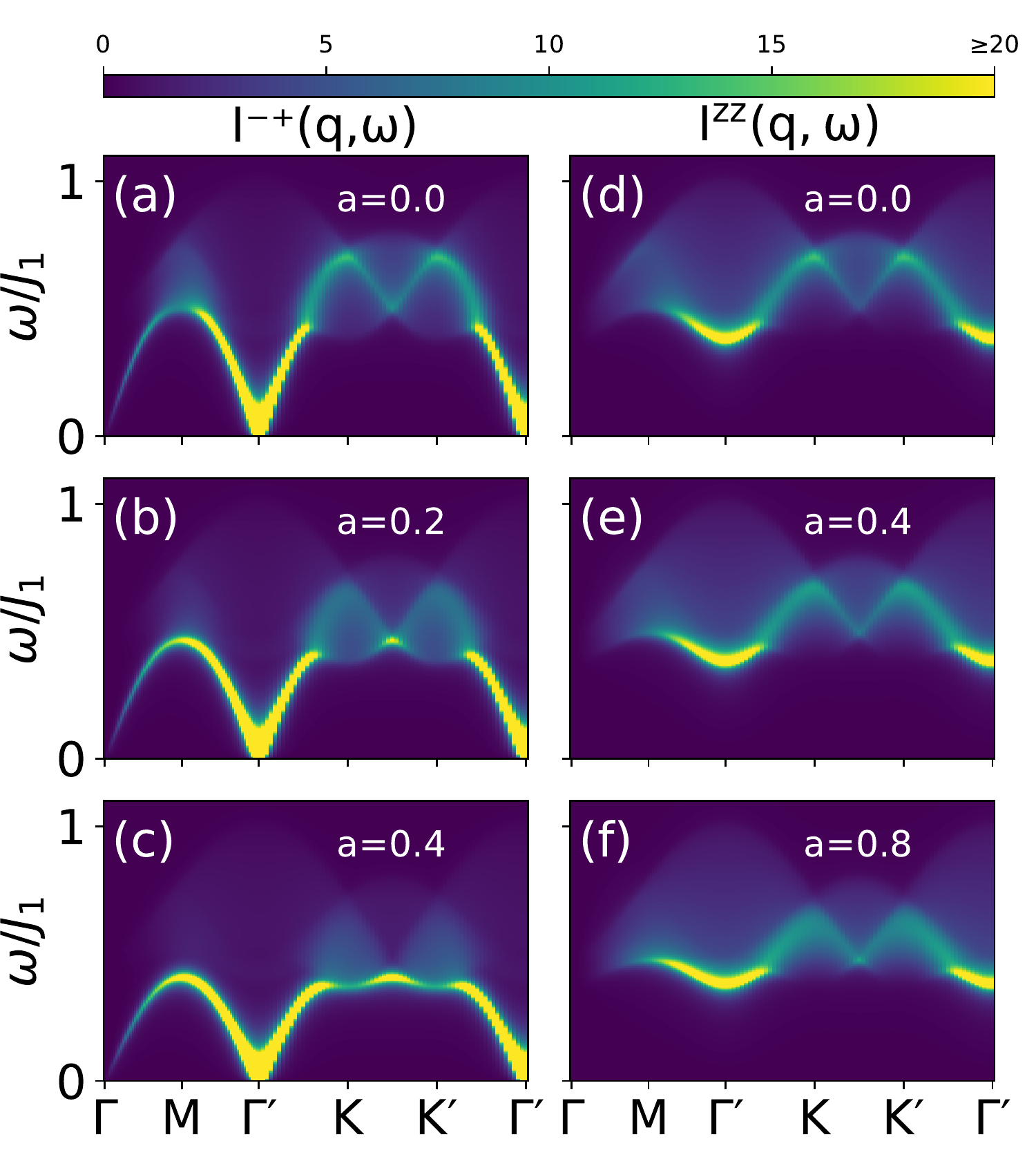}
\caption{(color online). Spectral functions of the (a)-(c) transverse and (d)-(f) longitudinal spin excitations along the $\Gamma$-$M$-$\Gamma'$-$K$-$K$-$\Gamma'$ path.}
\label{spectra}
\end{figure}

\par Now let's turn to the numerical results of spectra. In Figs.~\ref{spectra}(a)-(c), the transverse spectra of the AF phase with SU($2$) RVB fluctuations in the weak-frustration regime along the $\Gamma$-$M$-$\Gamma'$-$K$-$M$-$K'$-$\Gamma'$ path in the BZ [see Fig.~\ref{IPMC}(b)] are presented with three typical $a$ values. The spectra are composed of a dim broad spinon particle-hole continuum and a bright branch of magnons. The energies of the spinon pairs are $\omega_\mathbf{k}(\mathbf{q})=\varepsilon(\mathbf{k-q})+\varepsilon(\mathbf{k})$, thus, the continuum is gapped with the bottom lying at $\Gamma$/$K$. The magnons appear below the spinon continuum with a linear dispersion relation away from the gapless Goldstone mode at $\Gamma$/$\Gamma'$ as well as a divergent intensity at $\Gamma'$, which is in agreement with the SWT. However, as the magnons are excited to the high-energy region, the magnonic spectra exhibit distinct behaviors along the BZ boundary from $K$ to $M$, compared to the SWT predictions. When the Hubbard interaction is absent [Fig.~\ref{spectra}(a)], the bright magnons are scattered into the continuum as the dispersion ascends to $K$, and are completely absent along the $K$-$M$ line. As the Hubbard interaction sets in, the magnon dispersion is suppressed and the spectral weight is transferred to the lower boundary of the continuum. With a weak Hubbard interaction as in Fig.~\ref{spectra}(b), a bright mode comes out around $M$, but the spectrum at $K$ still remains as a dim continuum. With the further increase of the Hubbard interaction as in Figs.~\ref{spectra}(c), the transferred spectral weight at $K$ eventually concentrates on a roton-like mode with a local minimum in the dispersion. In Figs.~\ref{spectra}(d)-(f), the longitudinal spectra are also presented. From a spectral perspective, one would ascribe the bright mode, emerging adjoint to the bottom of the spinon continuum around $\Gamma'$ with gradually weakened spectral peaks approaching to the BZ boundary, as the well-defined collective longitudinal modes known as the Higgs modes. Different from the transverse spectra, the longitudinal ones are nearly independent of the Hubbard interaction strength, except that a bright mode would appear around $M$ when $a$ is extremely large [Fig.~\ref{spectra}(f)].

\begin{figure}
\centering
\includegraphics[width=0.40\textwidth]{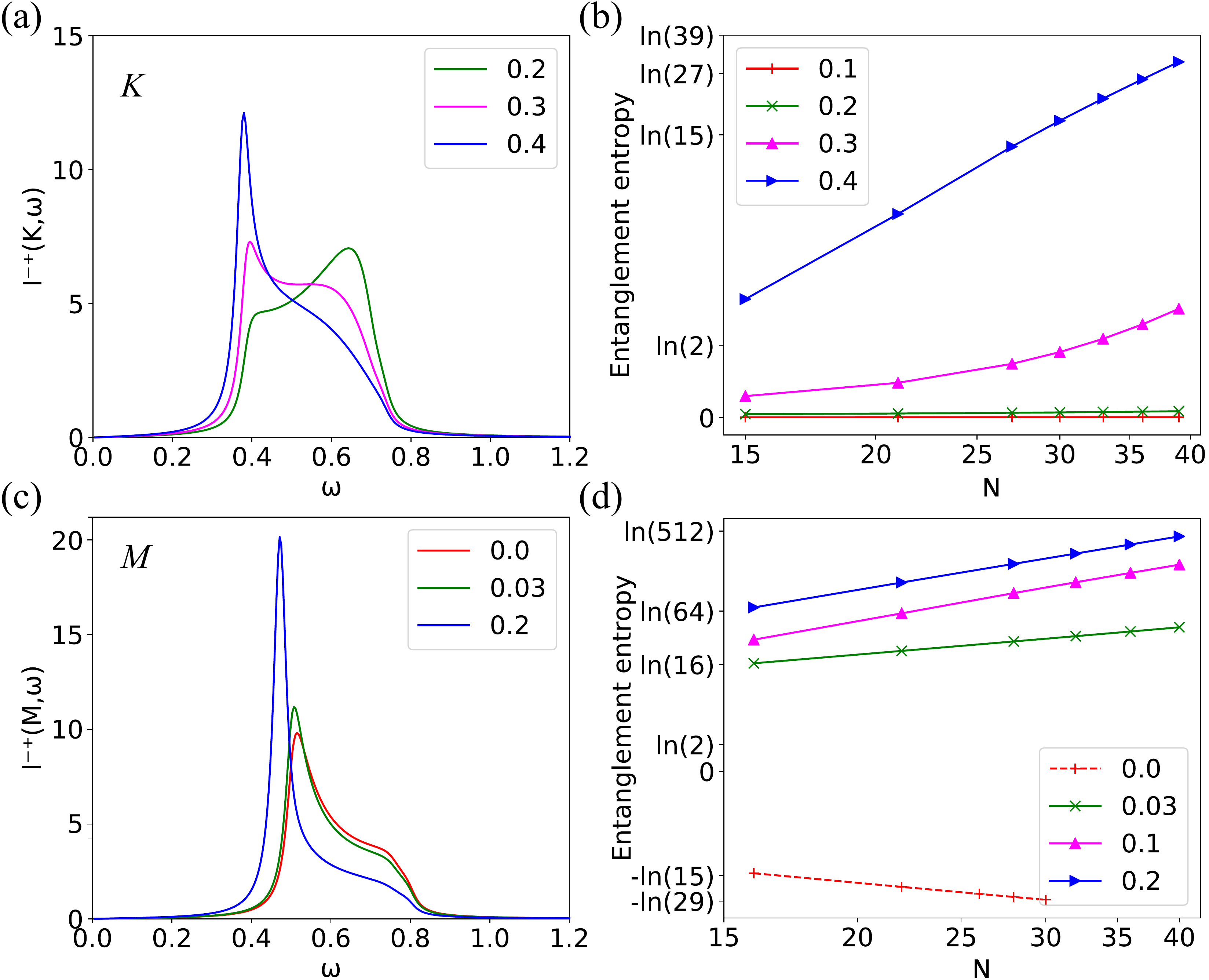}
\caption{(color online). On the left: transverse spectral functions at (a) $K$ and (b) $M$. On the right: scalings of the EEs of the lowest transverse modes at (c) $K$ and (d) $M$.}
\label{EET}
\end{figure}

\par
\begin{figure}
\centering
\includegraphics[width=0.40\textwidth]{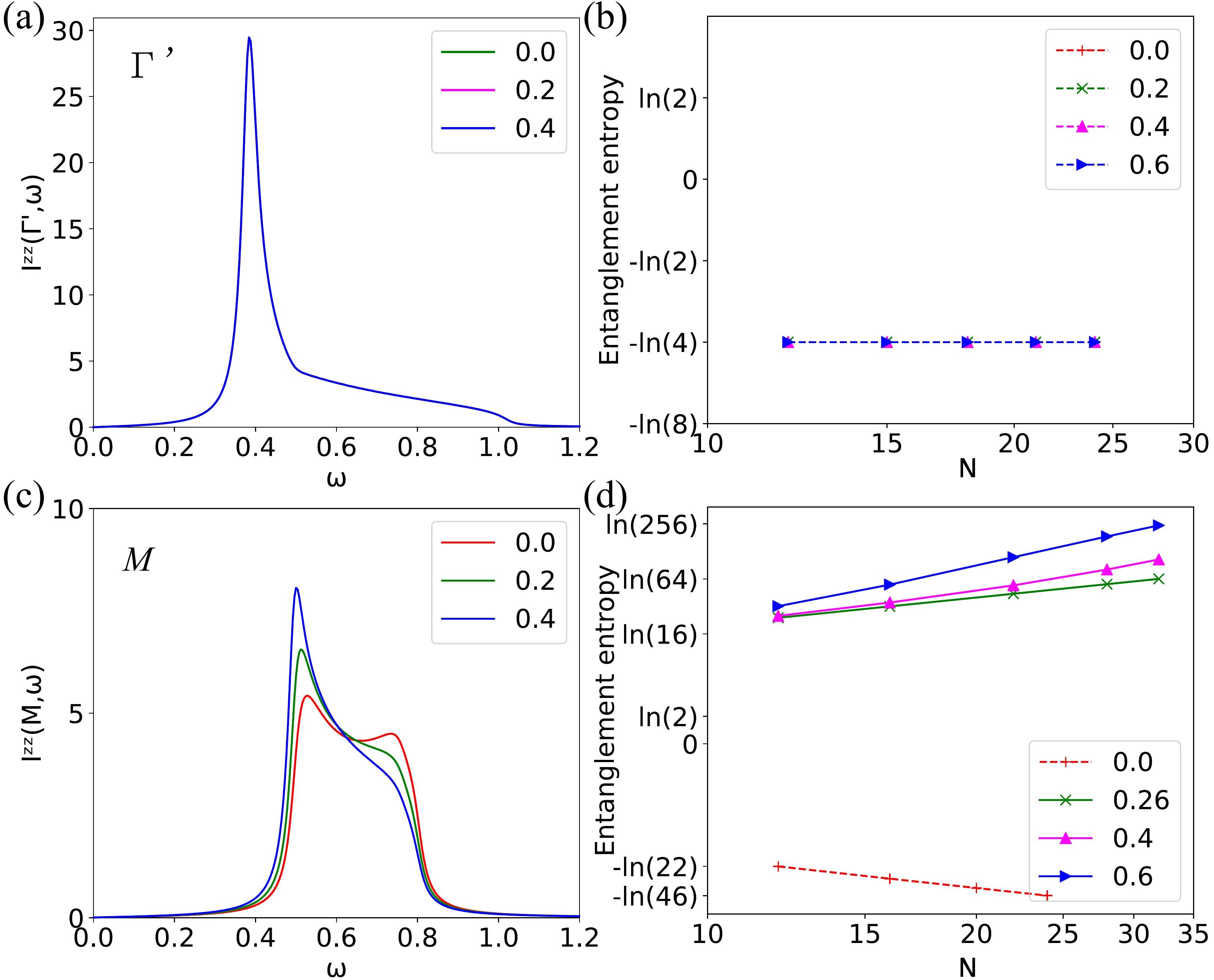}
\caption{(color online). On the left: longitudinal spectral functions at (a) $\Gamma'$ and (c) $M$. On the right: scalings of the EEs of the lowest longitudinal modes at (b) $\Gamma'$ and (d) $M$.}
\label{EEL}
\end{figure}

 To better understand the nature of the modes close to the continuum, we highlight the evolutions of the spectral functions with respect to $a$ for the transverse spin excitations at $M$/$K$ in Figs.~\ref{EET}(a)/(c), and for the longitudinal spin excitations at $\Gamma'$/$M$ in Figs.~\ref{EEL}(a)/(c). Basically, the determination of the magnons and spinons from the spectral perspective is based on the line shape of the spectral function. That is, the sharp Lorentzian quasiparticle peaks are identified as magnons, while the continua with a non-Lorentzian shape adhere to the Lorentzian peak are spinons. Thus, it is empirical. Here, we perform the $\alpha=2$ R\'{e}nyi entropy calculations with different system sizes and show the results for the quasiparticles with the lowest energies in Figs.~\ref{EET}(b)/(d) for the transverse modes at $K$/$M$, and in Figs.~\ref{EEL}(b)/(d) for the longitudinal modes at $\Gamma'$/$M$. There, $N$ is the number of the unit cells along either of the two translation vectors of the honeycomb lattice. Our EE analysis reveals the insufficiency of the spectral perspective and provides an quantitatively accurate judgement on the confinement and deconfinement of spinons, as will be elaborated below.

\par For the transverse modes at $K$ [Fig.~\ref{EET}(a)], as has been discussed above, the spectral weight transfers to the lower energy as the Hubbard interaction grows. Besides, the position of the spectral peak changes from the high-energy side to the low-energy side as $a$ grows across $\simeq0.3$, which corresponds to the emergence of a bright mode in the spectra and is a sign of the avoided fractionalization of the magnons into spinons in the strong interacting regime. This observation is verified by the EE results in Fig.~\ref{EET}(b). As is clearly seen, their EEs converge to constants when $a\lesssim0.3$ and diverge logarithmically when $a\gtrsim0.3$. Thus, there is a deconfinement-to-confinement (spinon-to-magnon) transition of the spin excitations across $\simeq0.3$. In this case, the spectral analysis coincides with the EE analysis. However, for the transverse modes at $M$, as can be seen in Fig.~\ref{EET}(c), no qualitative change of the line shapes of the spectral functions could be observed when the Hubbard interaction increases, especially from $a=0$ to $a\simeq 0.03$. It is hard to determine whether magnons are fractionalized into spinons at this point from the spectral perspective. To resolve this dilemma, we have to resort to the EE analysis. As shown in Fig.~\ref{EET}(d), the EE of the lowest-energy mode at M is logarithmically divergent with the system size when the Hubbard interaction exceeds a critical value ($a\simeq 0.03$). Such a specific scaling behavior implies the formation of stable magnons with $a\gtrsim 0.03$. On the other hand, when $a\simeq0$, the EE scales proportional to $-{\rm ln}(N-1)$ \cite{DegenerateEE}. This is related to the $N-1$ fold degeneracy of the lowest-energy spinon particle-hole pairs. For a single mode in the degenerate manifold, the averaged EE converges to a constant. Thus, the spin excitations at $M$ with $a\simeq0$ are deconfined spinons.

\par For the longitudinal modes at $\Gamma'$, regardless of the Hubbard interaction strength the spin excitations seem to be well-defined Higgs modes at first sight. After all, they exhibit quite sharp peaks in the spectral function [see Fig.~\ref{EEL}(a)]. However, the convergence of the EEs in Fig.~\ref{EEL}(b) rebuts this physical picture implied by the spectral function, and $S(\mathbf{q})=-{\rm ln}4$ denotes four degenerate free spinon particle-hole pair excitations. At $M$, the case for the longitudinal mode is similar to that of the transverse mode, where the confinement and deconfinement of spinons cannot be determined by the spectral functions due to their continuous evolution with $a$ [see Fig.~\ref{EEL}(c)]. By the evaluation of the EEs as in Fig.~\ref{EEL}(d), it can be concluded that a stable Higgs mode emerges as $a\gtrsim0.26$.

\par In summary, we investigate the spin excitation spectra of the antiferromagnet in the honeycomb lattice in the weak-frustration regime and introduce an EE analysis to identify the confinement and deconfinement of the spinons. The spectra exhibit anomalous behaviors in the high-energy region where the collective modes along the BZ boundary are strongly renormalized by the spinon continuum. With the help of the EE analysis, we elaborate quantitatively the deconfinement-to-confinement transition of spinons in the anomalous spectra with the increase of the Hubbard interaction, indicating the avoided fractionalization of magnons in the strong interaction regime. Meanwhile, the Higgs mode at the $\Gamma'$ point is always deconfined as spinons regardless of the Hubbard interaction strength, although it appears as a sharp well-defined quasiparticle from the spectral perspective.

\par The anomalous continuum behavior of the spin excitations in the recent experiments of YbCl$_3$ and YbBr$_3$ \cite{PhysRevB.102.014427, Wessler2020, PhysRevB.100.180406, Sala2021} is close to the $a=0.2\sim0.3$ case in this paper, where the spectra peak at $K$ is more broadening than that at $M$, and is not pushed down to be a well-defined roton-like collective mode. But the experiments do not show a definite bottom of the continuum at $K$. Thus, the effective models of the two compounds remain to be verified. Moreover, the fluctuations induced by the frustrations may not be limited to the SU($2$) RVB type. Investigations beyond the mean-field level is needed to determine whether a Z$_2$ spin-liquid type is more favourable.

\par It is noted that there exists another scenario to account for the anomalous spectra in the high-energy region, which is attributed to the magnon damping \cite{PhysRevLett.118.107202, PhysRevLett.119.157201, PhysRevLett.120.087201, PhysRevLett.86.528,PhysRevB.79.144416,PhysRevB.88.094407, SciPostPhys.4.1.001, Sala2021} instead of the deconfinement of spinons. In principle, the EEs for damped magnons and deconfined spinons should have different scaling behaviors. The EE analysis could be generalized to more unbiased methods in future works to help settle down this debate.

\begin{acknowledgments}
\par This work was supported by the National Natural Science Foundation of China (Grants No. 11904170, 11674158, 11774152, and 12004191), the Natural Science Foundation of Jiangsu Province, China (Grant No. BK20190436 and BK20200738), National Key Projects for Research and Development of China (Grant No. 2016YFA0300401), and Doctoral Program of Innovation and Entrepreneurship in Jiangsu Province.
\end{acknowledgments}

\bibliography{reference}

\begin{thebibliography}{51}
\bibitem [{Deg()}]{DegenerateEE}%
  \BibitemOpen
  \href@noop {} {\bibinfo  {journal} {Here an extra sum over the degenerate
  states is applied in Eq.~\ref{EE} since in numerics the EE for a single mode
  in the degenerate manifold is not stable due to the degeneracy. Thus, the
  sumed EEs $S\propto ln D$ if the excitations are free spinons, where $D$
  is the degeneracy.}\ }\BibitemShut {NoStop}%
\end{thebibliography}

\end{document}